\documentclass{emulateapj}
\slugcomment{Version 10/12/17.  Preprint: FERMILAB-PUB-17-452-AE . DES-2017-0279}
\usepackage{graphicx}
\graphicspath{ {} }
\usepackage{pgf,tikz}
\usepackage{mathrsfs}
\usepackage{comment}
\usepackage{fancyhdr}

\usepackage{amsmath}
\usetikzlibrary{arrows}
\usepackage{pgffor}
\usepackage{xcolor}
\usepackage{lineno}
\usepackage{textcomp}
\usepackage{makecell}

\usepackage{eso-pic}% http://ctan.org/pkg/eso-pic

\newcommand{\DKN}{40}  % distance, in Mpc
\newcommand{\zKN}{0.0098}  % redshift

\newcommand{\NSURVEY}{11}

\newcommand{\RateUnit}{\rm{Gpc}^{-3}/\rm{year}}
\newcommand{\KNRATE}{{10^3}}
\newcommand{\zmaxWFIRST}{{0.8}}
\newcommand{\zmaxLSSTWIDE}{{0.25}}
\newcommand{\zmaxZTF}{{0.04}}

\newcommand{\NKNDES}{{0.26}}
\newcommand{\NYDES}{{5}}

\newcommand{\NKNSNLS}{{0.11}}
\newcommand{\NYSNLS}{{4}}

\newcommand{\NKNSDSS}{{0.13}}
\newcommand{\NYSDSS}{{2}}

\newcommand{\NKNPS}{{0.22}}
\newcommand{\NYPS}{{4}}

\newcommand{\NKNATLAS}{{8.3}}
\newcommand{\NYATLAS}{{5}}

\newcommand{\NKNZTF}{10.6}
\newcommand{\NYZTF}{{5}}

\newcommand{\NKNSMT}{0.001}
\newcommand{\NYSMT}{{5}}

\newcommand{\NKNLSSTWIDE}{{69}}  % number of KN passing cuts, just 2 sig digits
\newcommand{\NYLSSTWIDE}{{10}}  % number of KN passing cuts

\newcommand{\NKNLSSTDEEP}{{5.5}}  % number of KN passing cuts
\newcommand{\NYLSSTDEEP}{{10}}  % number of KN passing cuts

\newcommand{\NKNWFIRST}{{16.0}}  % number of KN passing cuts
\newcommand{\NYWFIRST}{{2}}  % number of KN passing cuts

\begin{document}

\title{How Many Kilonovae Can Be Found in Past, Present, and Future Survey Datasets?}

\author{
D.~Scolnic\altaffilmark{1},
R.~Kessler\altaffilmark{1},
D.~Brout\altaffilmark{2},
P.~S.~Cowperthwaite\altaffilmark{3},
M.~Soares-Santos\altaffilmark{4,5},
J.~Annis\altaffilmark{4},
K.~Herner\altaffilmark{4},
H.-Y.~Chen\altaffilmark{1},
M.~Sako\altaffilmark{2},
Z.~Doctor\altaffilmark{1},
R.~E.~Butler\altaffilmark{6,7},
A.~Palmese\altaffilmark{8},
H.~T.~Diehl\altaffilmark{4},
J.~Frieman\altaffilmark{4,1},
D.~E.~Holz\altaffilmark{1,9,10,9},
E.~Berger\altaffilmark{3},
R.~Chornock\altaffilmark{11},
V.~A.~Villar\altaffilmark{3},
M.~Nicholl\altaffilmark{3},
R.~Biswas\altaffilmark{12,13},
R.~Hounsell\altaffilmark{14,15},
R.~J.~Foley\altaffilmark{14},
J.~Metzger\altaffilmark{1},
A.~Rest\altaffilmark{16,17},
J.~Garc\'ia-Bellido\altaffilmark{18},
A.~M\"oller\altaffilmark{19,20},
P.~Nugent\altaffilmark{21},
T.~M.~C.~Abbott\altaffilmark{22},
F.~B.~Abdalla\altaffilmark{8,23},
S.~Allam\altaffilmark{4},
K.~Bechtol\altaffilmark{24},
A.~Benoit-L{\'e}vy\altaffilmark{25,8,26},
E.~Bertin\altaffilmark{25,26},
D.~Brooks\altaffilmark{8},
E.~Buckley-Geer\altaffilmark{4},
A.~Carnero~Rosell\altaffilmark{27,28},
M.~Carrasco~Kind\altaffilmark{29,30},
J.~Carretero\altaffilmark{31},
F.~J.~Castander\altaffilmark{32},
C.~E.~Cunha\altaffilmark{33},
C.~B.~D'Andrea\altaffilmark{2},
L.~N.~da Costa\altaffilmark{27,28},
C.~Davis\altaffilmark{33},
P.~Doel\altaffilmark{8},
A.~Drlica-Wagner\altaffilmark{4},
T.~F.~Eifler\altaffilmark{34,35},
B.~Flaugher\altaffilmark{4},
P.~Fosalba\altaffilmark{32},
E.~Gaztanaga\altaffilmark{32},
D.~W.~Gerdes\altaffilmark{36,37},
D.~Gruen\altaffilmark{33,38},
R.~A.~Gruendl\altaffilmark{29,30},
J.~Gschwend\altaffilmark{27,28},
G.~Gutierrez\altaffilmark{4},
W.~G.~Hartley\altaffilmark{8,39},
K.~Honscheid\altaffilmark{40,41},
D.~J.~James\altaffilmark{42},
M.~W.~G.~Johnson\altaffilmark{30},
M.~D.~Johnson\altaffilmark{30},
E.~Krause\altaffilmark{33},
K.~Kuehn\altaffilmark{43},
S.~Kuhlmann\altaffilmark{44},
O.~Lahav\altaffilmark{8},
T.~S.~Li\altaffilmark{4},
M.~Lima\altaffilmark{45,27},
M.~A.~G.~Maia\altaffilmark{27,28},
M.~March\altaffilmark{2},
J.~L.~Marshall\altaffilmark{46},
F.~Menanteau\altaffilmark{29,30},
R.~Miquel\altaffilmark{47,31},
E.~Neilsen\altaffilmark{4},
A.~A.~Plazas\altaffilmark{35},
E.~Sanchez\altaffilmark{48},
V.~Scarpine\altaffilmark{4},
M.~Schubnell\altaffilmark{37},
I.~Sevilla-Noarbe\altaffilmark{48},
M.~Smith\altaffilmark{49},
R.~C.~Smith\altaffilmark{22},
F.~Sobreira\altaffilmark{50,27},
E.~Suchyta\altaffilmark{51},
M.~E.~C.~Swanson\altaffilmark{30},
G.~Tarle\altaffilmark{37},
R.~C.~Thomas\altaffilmark{21},
D.~L.~Tucker\altaffilmark{4},
A.~R.~Walker\altaffilmark{22}
\\ \vspace{0.2cm} (DES Collaboration) \\
}

\affil{$^{1}$ Kavli Institute for Cosmological Physics, University of Chicago, Chicago, IL 60637, USA}
\affil{$^{2}$ Department of Physics and Astronomy, University of Pennsylvania, Philadelphia, PA 19104, USA}
\affil{$^{3}$ Harvard-Smithsonian Center for Astrophysics, 60 Garden Street, Cambridge, MA 02138, USA}
\affil{$^{4}$ Fermi National Accelerator Laboratory, P. O. Box 500, Batavia, IL 60510, USA}
\affil{$^{5}$ Department of Physics, Brandeis University, Waltham, MA 02453, USA}
\affil{$^{6}$ Department of Astronomy, Indiana University, 727 E. Third Street, Bloomington, IN 47405, USA}
\affil{$^{7}$ Fermi National Accelerator Laboratory, P.O. Box 500, Batavia, IL 60510, USA}
\affil{$^{8}$ Department of Physics \& Astronomy, University College London, Gower Street, London, WC1E 6BT, UK}
\affil{$^{9}$ Department of Physics, University of Chicago, Chicago, IL 60637, USA}
\affil{$^{10}$ Enrico Fermi Institute, University of Chicago, Chicago, IL 60637, USA}
\affil{$^{11}$ Astrophysical Institute, Department of Physics and Astronomy, 251B Clippinger Lab, Ohio University, Athens, OH 45701, USA}
\affil{$^{12}$ The eScience Institute and the Department of Astronomy, University of Washington, Seattle, WA 98195,USA}
\affil{$^{13}$ Oskar Klein Centre, Department of Physics, Stockholm University, SE 106 91 Stockholm, Sweden}
\affil{$^{14}$ Department of Astronomy and Astrophysics, University of California Santa Cruz, 1156 High St., Santa Cruz, CA 95064, USA}
\affil{$^{15}$ Department of Astronomy, University of Illinois Urbana Champaign, 1002 W Green St., Urbana, IL, 61801, USA}
\affil{$^{16}$ Space Telescope Science Institute, 3700 San Martin Dr., Baltimore, MD 21218, USA}
\affil{$^{17}$ Department of Physics and Astronomy, The Johns Hopkins University, 3400N. Charles St., Baltimore, MD 21218, USA}
\affil{$^{18}$ Instituto de Fisica Teorica UAM/CSIC, Universidad Autonoma de Madrid, 28049 Madrid, Spain}
\affil{$^{19}$ ARC Centre of Excellence for All-sky Astrophysics (CAASTRO)}
\affil{$^{20}$ The Research School of Astronomy and Astrophysics, Australian National University, ACT 2601, Australia}
\affil{$^{21}$ Lawrence Berkeley National Laboratory, 1 Cyclotron Road, Berkeley, CA 94720, USA}
\affil{$^{22}$ Cerro Tololo Inter-American Observatory, National Optical Astronomy Observatory, Casilla 603, La Serena, Chile}
\affil{$^{23}$ Department of Physics and Electronics, Rhodes University, PO Box 94, Grahamstown, 6140, South Africa}
\affil{$^{24}$ LSST, 933 North Cherry Avenue, Tucson, AZ 85721, USA}
\affil{$^{25}$ CNRS, UMR 7095, Institut d'Astrophysique de Paris, F-75014, Paris, France}
\affil{$^{26}$ Sorbonne Universit\'es, UPMC Univ Paris 06, UMR 7095, Institut d'Astrophysique de Paris, F-75014, Paris, France}
\affil{$^{27}$ Laborat\'orio Interinstitucional de e-Astronomia - LIneA, Rua Gal. Jos\'e Cristino 77, Rio de Janeiro, RJ - 20921-400, Brazil}
\affil{$^{28}$ Observat\'orio Nacional, Rua Gal. Jos\'e Cristino 77, Rio de Janeiro, RJ - 20921-400, Brazil}
\affil{$^{29}$ Department of Astronomy, University of Illinois, 1002 W. Green Street, Urbana, IL 61801, USA}
\affil{$^{30}$ National Center for Supercomputing Applications, 1205 West Clark St., Urbana, IL 61801, USA}
\affil{$^{31}$ Institut de F\'{\i}sica d'Altes Energies (IFAE), The Barcelona Institute of Science and Technology, Campus UAB, 08193 Bellaterra (Barcelona) Spain}
\affil{$^{32}$ Institute of Space Sciences, IEEC-CSIC, Campus UAB, Carrer de Can Magrans, s/n,  08193 Barcelona, Spain}
\affil{$^{33}$ Kavli Institute for Particle Astrophysics \& Cosmology, P. O. Box 2450, Stanford University, Stanford, CA 94305, USA}
\affil{$^{34}$ Department of Physics, California Institute of Technology, Pasadena, CA 91125, USA}
\affil{$^{35}$ Jet Propulsion Laboratory, California Institute of Technology, 4800 Oak Grove Dr., Pasadena, CA 91109, USA}
\affil{$^{36}$ Department of Astronomy, University of Michigan, Ann Arbor, MI 48109, USA}
\affil{$^{37}$ Department of Physics, University of Michigan, Ann Arbor, MI 48109, USA}
\affil{$^{38}$ SLAC National Accelerator Laboratory, Menlo Park, CA 94025, USA}
\affil{$^{39}$ Department of Physics, ETH Zurich, Wolfgang-Pauli-Strasse 16, CH-8093 Zurich, Switzerland}
\affil{$^{40}$ Center for Cosmology and Astro-Particle Physics, The Ohio State University, Columbus, OH 43210, USA}
\affil{$^{41}$ Department of Physics, The Ohio State University, Columbus, OH 43210, USA}
\affil{$^{42}$ Astronomy Department, University of Washington, Box 351580, Seattle, WA 98195, USA}
\affil{$^{43}$ Australian Astronomical Observatory, North Ryde, NSW 2113, Australia}
\affil{$^{44}$ Argonne National Laboratory, 9700 South Cass Avenue, Lemont, IL 60439, USA}
\affil{$^{45}$ Departamento de F\'isica Matem\'atica, Instituto de F\'isica, Universidade de S\~ao Paulo, CP 66318, S\~ao Paulo, SP, 05314-970, Brazil}
\affil{$^{46}$ George P. and Cynthia Woods Mitchell Institute for Fundamental Physics and Astronomy, and Department of Physics and Astronomy, Texas A\&M University, College Station, TX 77843,  USA}
\affil{$^{47}$ Instituci\'o Catalana de Recerca i Estudis Avan\c{c}ats, E-08010 Barcelona, Spain}
\affil{$^{48}$ Centro de Investigaciones Energ\'eticas, Medioambientales y Tecnol\'ogicas (CIEMAT), Madrid, Spain}
\affil{$^{49}$ School of Physics and Astronomy, University of Southampton,  Southampton, SO17 1BJ, UK}
\affil{$^{50}$ Instituto de F\'isica Gleb Wataghin, Universidade Estadual de Campinas, 13083-859, Campinas, SP, Brazil}
\affil{$^{51}$ Computer Science and Mathematics Division, Oak Ridge National Laboratory, Oak Ridge, TN 37831}

%\pagestyle{headings}
%\pagestyle{fancy}
%\rhead{WFIRST BLA BLA}

\begin{abstract}
The discovery of a kilonova (KN) associated with the Advanced LIGO (aLIGO)/Virgo event GW170817 opens up new avenues of multi-messenger astrophysics.  
Here, using realistic simulations, we provide estimates of the number of KNe that could be found in data from past, present and future surveys without a gravitational-wave trigger.  
For the simulation, we construct a spectral time-series model based on the DES-GW multi-band light-curve from the single known KN event, and we use an average of BNS rates from past studies of
${\KNRATE}~\RateUnit$, consistent with the $1$ event found so far.
Examining past and current datasets from transient surveys, the number of KNe we expect to find for 
ASAS-SN, SDSS, PS1, SNLS, DES, and SMT
is between 0 and $0.3$.  
We predict the number of detections per future survey to be: {\NKNATLAS} from ATLAS, {\NKNZTF} from ZTF, {\NKNLSSTDEEP}/{\NKNLSSTWIDE} from LSST (the Deep Drilling / Wide Fast Deep), and {\NKNWFIRST} from WFIRST. The maximum redshift of KNe discovered for each survey is
$z={\zmaxWFIRST}$ for WFIRST, $z={\zmaxLSSTWIDE}$ for LSST and $z={\zmaxZTF}$ for ZTF and ATLAS.  
This maximum redshift for WFIRST is well beyond the sensitivity of aLIGO and some future GW missions.   
For the LSST survey, we also provide contamination estimates from Type Ia and Core-collapse supernovae:
after light-curve and template-matching requirements, we estimate a background of just 2 events.
More broadly, we stress that future transient surveys should consider how to optimize their search strategies to improve their detection efficiency, and to consider similar analyses for GW follow-up programs.

\end{abstract}


\section{Introduction}
The first detection by aLIGO/Virgo of a GW signal from a binary neutron star coalescence 
\citep{LigoGCN1,LigoGCN2} and the identification of the optical 
counterpart \citep{CoulterGCN,SoaresGCN,ValentiGCN} marks the beginning of an 
exciting era of joint electromagnetic (EM) and gravitational-wave (GW) studies.  
Optical counterparts from the mergers of a binary containing a neutron star are 
called `kilonova' (hereafter KN, see \citealp{Metzger17} for a review and references therein).  
Theoretical studies predict that outflows of neutron-rich material during the 
merger enable r-process nucleosynthesis, and 
that the decay of these r-process elements results in isotropic thermal emission. 
As KN events result in visible transients in the optical and infrared,
with time scales of hours to days, \cite{Metzger:2011bv} have predicted 
that nearby KNe may be bright enough to find with modern optical telescopes.  
These predictions have been confirmed.

Optical observations of KNe can constrain theories about neutron star mergers, 
in particular identifying them as the progenitors of short Gamma Ray Bursts (GRBs).  
These events can also be used to measure the current expansion rate of the universe 
if there is a GW signal and the
associated host galaxy redshift can be measured (e.g., \citealp{Schutz,Dalal}).
Additionally, untriggered KN discoveries in the optical would help LIGO re-evaluate marginal signals and improve their detection algorithms.
%For example, \cite{Dalal} and \cite{Nissanke} showed that
%with large numbers (20-50) of KN events out to $z\sim0.1$, percent-level $H_0$ measurements can be determined.  

To date, there have been a small number of inconclusive KN detections 
(e.g., \citealp{Tanvir, Berger:2013wna,Jin2016}), none of which were triggered by a transient survey.
With an optical counterpart of a GW event having been discovered, 
this event can be used to estimate the volumetric rate of KN events.  
 Making the simplistic assumptions that 
 all KN events are the same and that the volumetric rate is constant with redshift, we can predict how many of these events can be found in past, present, and future surveys.  
This is a follow-up of the work by \cite{Doctor17} who considered a wide range of KN models and examined 
2 seasons of data from the Dark Energy Survey Supernova Program (DES-SN).
Here, we examine a single model, but a wide range of surveys.  
Other studies (e.g., \citealp{Rosswog}) have considered the detectability of KNe with future surveys based on estimated search depths, but here we consider depth, cadence, and area of the surveys using realistic observation libraries.

 In this paper, we use simulations to assess the capabilities of photometric surveys to discover KNe without a GW trigger.  
This is different from the follow-up mode for GW170817 and for past EM searches \citep{GW151226,Annis,SoaresSantos,CowperthwaiteFollow}
that followed a GW trigger from LIGO (\citealt{ligo1}, \citealt{ligo2}, \citealt{ligo3}). 
Over the last decade there has been a large effort in predicting biases for SNIa distance measurements that are used as cosmological probes 
(e.g., \citealp{Scolnic17c}), and this effort has resulted in increasingly realistic simulations.  
The SNANA \citep{SNANA} software used for these studies has  
been applied to many cases beyond Type Ia supernovae, 
including core-collapse SNe, superluminous SNe and kilonovae \citep{Doctor17}.
All simulation and analysis tools used here are publicly 
available.\footnote{snana.uchicago.edu}

In Section 2 we briefly review the KN discovery, and use companion works to model the light curve and KN rate.  
In Section 3, we describe {\NSURVEY}  optical surveys  and our simulation methods.
Results are presented in Section 4, along with estimates of the background contamination  from SNe. 
Finally, in Section 5, we discuss future analyses  to optimize these surveys, and present conclusions in Section 6.

\section{The Optical Counterpart to LV G298048}
\subsection{Discovery of Counterpart}
Just over 11 hours after the aLIGO trigger \citep{LigoGCN1,LigoGCN2},
the optical counterpart was found 
\citep{CoulterGCN, SoaresGCN,ValentiGCN}.  
The counterpart was identified as a point source
located near NGC 4993.
This galaxy is \DKN\ Mpc away, with redshift $z=\zKN$  \citep{Kourkchi17}.  In a companion paper by \cite{SoaresGCN}, we use our DECam \citep{Flaugher15} search data to show the likelihood that the transient is in fact directly 
connected to the GW event is $>99\%$.  We therefore rely on this event for our analysis.  
\begin{figure*}
\begin{center}
\hspace*{-0.2in} 
\scalebox{1.}
{\includegraphics[width=1.0\textwidth]{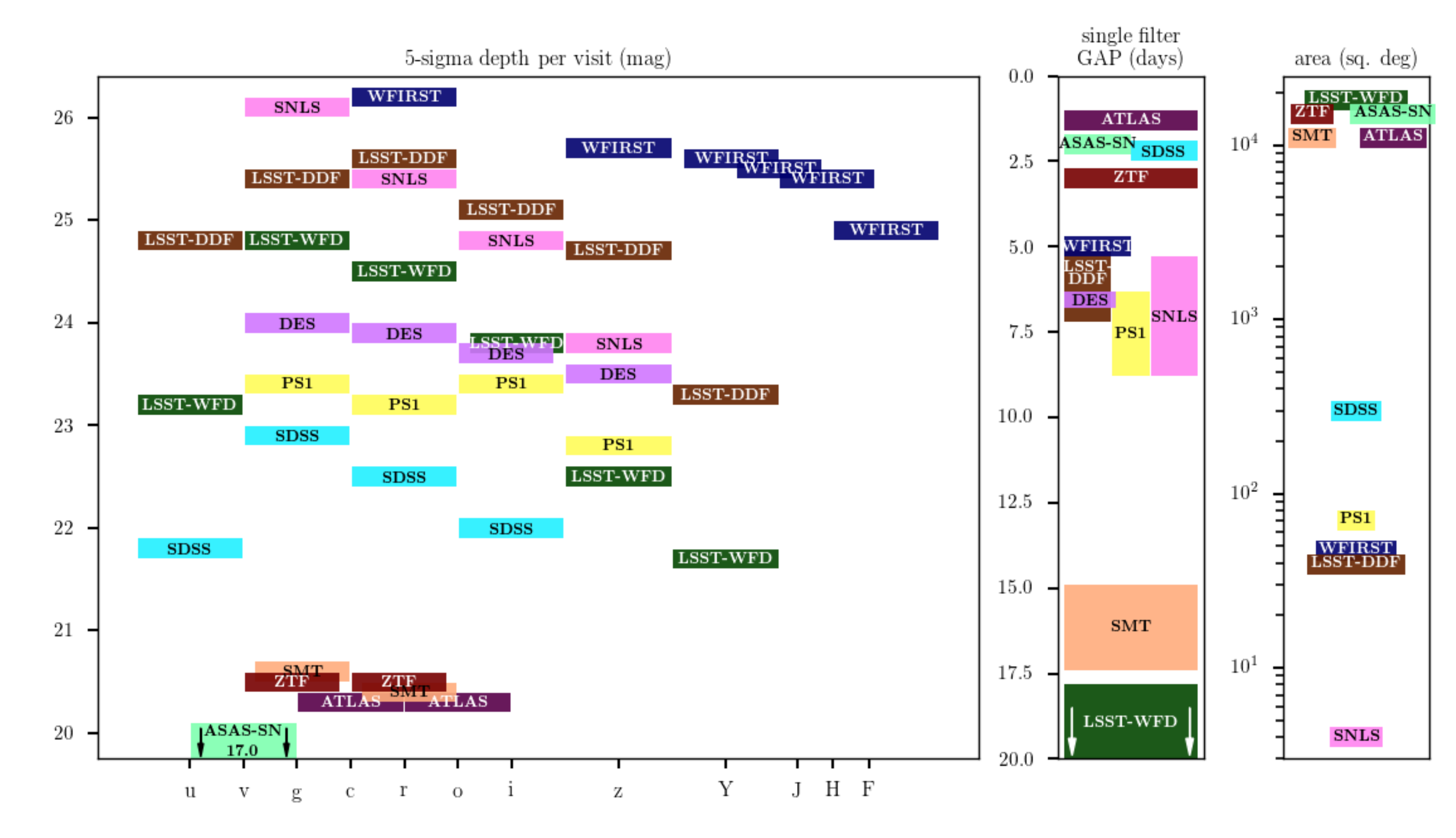}}
\caption{Display of key characteristics for \textit{transient surveys} used in our analysis. 
Left panel shows the depth per night per filter. 
Middle panel shows the mean gap between repeat observations in a single filter. 
Right panel shows the survey area covered each observing-year.}
\label{fig:Omh}
\end{center}
\end{figure*}

\subsection{KN Lightcurve and Modeling}
%For our simulations, we need a spectral model with UV to NIR coverage over the duration of the lightcurve.  
 We model UV to NIR to allow for a broad range of analysis. While \cite{Doctor17} showed that a NIR model is sufficient (e.g., $i$ and $z$ bands) for estimating KN detections, the bluer bands can be used to reject backgrounds from supernovae.
Our KN model is determined using $ugrizYHK$ photometry from the DES-GW papers 
\citep{SoaresGCN,CowperthwaiteLC}.  
To build a spectral model for simulations, we take the smoothed lightcurves from \cite{CowperthwaiteLC} in $ugrizYHK$ and ``mangle'' (Hsiao et al. 2007) a spectral time series to match
the observed photometry.  The mangling uses wavelength-dependent
splines with knots at the effective wavelengths of the 8 photometric filters.
Our model has peak $i-z\sim0.0$ and fades roughly 5 magnitudes over 7 days, in agreement with the  data.

\cite{CowperthwaiteLC} shows that this KN includes both a blue and red component, resulting in early time colors that are
bluer than most models which predict  $i-z \sim 1$~mag 
(e.g., \citealp{Barnes2016}). While our observed KN may not reflect the general population, we do not attempt to speculate about the population properties.

%any further than the one object we have.  
%This is analyzed further in the discussion.

%Cowperthwaite et al. explains that models with two components for lanthanide-poor and lanthanide-rich ejecta provide a good fit for the data and result in a blue and red component.  This explanation differs from most predicted models with redder colors for the KNe.  

\subsection{Estimate of Volumetric Rate}
We use a constant volumetric KN rate of ${\KNRATE}~\RateUnit$ as a conservative estimate  based on a compilation of rates by \cite{Abbot2016}.  This estimate is consistent with the fact that the LIGO O1 upper limit is 
$1.2\times10^4~\RateUnit$~\citep{Abbot2016}, and O2 surveyed $\sim10$ times more volume than O1, suggesting a rate of $\sim10^3~\RateUnit$.  
Furthermore, this rate is broadly consistent with the aLIGO search time ($<2$ years) and 
search volume $\sim (100~{\rm Mpc})^3$.
%fact that there has been 1 GW event with a KN from $<2$ years of aLIGO 
%searching and the event  at 40-50 Mpc was detected well above the sensitivity limits. 

%that includes measurements by \cite{Kim,Fong,Siellez,Coward,Petrillo,Jin,Vangioni,deMink,Dominik}.  

% as our nominal rate. 
%Importantly, this rate can be scaled up or down either in our simulations or from the results of our simulations, so is meant to be used a benchmark and can be updated with future rate constraints from LIGO.

\section{Simulation of Transient Surveys}
%\begin{comment}

\begin{table*}
\centering
\caption{
{ Summary Information for Each Survey.}
%The cadence is the mean duration between return visits in each filter.  
%The Area is the total amount of sky area covered per year.  
%The duration is the total number of years per survey.  
%The citation given is for where observation history or characteristics for each survey is described. 
}
\begin{tabular}{l|llllll}
\hline 
Survey & Filters & Depths & 
   Cadences\tablenotemark{a} & 
    Area\tablenotemark{b} & 
    Duration\tablenotemark{c} & 
    Citation\tablenotemark{d}  \\
~ & ~ & ~ [$5\sigma$ mag]& [Days] & [Sq. Deg] & [Years] & ~ \\
\hline
SDSS & $ugriz$ & $21.8,22.9,22.5,22.$ & $2.2,2.2,2.2,2.2$ & 300 & 2 & \cite{Frieman08} \\
SNLS & $griz$ & $26.1,25.4,24.8,23.8$ & $8.8,6.3,5.3,8.5$ & 4 & 5 & \cite{Astier06} \\
PS1  & $griz$ & $23.4,23.2,23.4,22.8$ & $8.8,8.7,8.2,6.3$ & 70 & 4 & \cite{Scolnic14} \\
DES & $griz$ & $24.0,23.9,23.7,23.5$ & $6.8,6.4,6.3,6.5$ & 27 & 5 & \cite{Kessler15} \\
ASAS-SN & $V$ & $17.5$ & $2$ & 15000 & 5 & \cite{Shappee14}  \\
SMT & $gr$ & $20.6,20.4$ & $17.4,14.9$ & 11000 & 5 & \cite{Scalzo17} \\
ATLAS & $co$ & $20.3,20.3$ & $1.3,1.3$ & 11000 & 5 & \cite{Tonry11} \\
ZTF & $gr$ & $20.5,20.5$ & $3.0,3.0$ & 15000 & 5 & \cite{ztf} \\
LSST DDF & $ugrizy$ & $24.8,25.4,25.6,25.1,24.7,23.3$ & 
%    $5.1,5.9,7.1,7,7.1,7.2$  & 
       $5, 6, 7, 7, 7, 7$  &    % round to nearest day to fit table width on page
      40 & 10 & \cite{LSST09} \\
LSST WFD & $ugrizy$ & $23.2,24.8,24.5,23.8,22.5,21.7$ & 
%         $30.0,34.9,17.8,18.6,21.0,17.8$ & 
          $30, 35, 18, 19, 21, 18$ & 
          18000 & 10 & \cite{LSST09} \\
WFIRST & $RZYJHF$ & $26.2,25.7,25.6,25.5,25.4,24.9$ & $5,5,5,5,5$ & 45 & 2 & \cite{Hounsell17} \\
\hline
\end{tabular}
\tablenotetext{1}{Mean duration between return visits in each filter.}
\tablenotetext{2}{Total amount of sky area covered per year.}
\tablenotetext{3}{Total number of years per survey.}
\tablenotetext{4}{Describes observation history or characteristics.}
\label{tab:hvals}
\end{table*}
For this analysis, we have selected large surveys with the following criteria:
they operate as rolling searches, and have (or expect to have) discovered at least 100 SNe.
The compilation of surveys is listed in Table~\ref{tab:hvals} and includes that from
The Sloan Digital Sky Survey-II (SDSS, \citealp{Frieman08}),  
Panoramic Survey Telescope and Rapid Response System (PS1, \citealp{Kaiser10},)
Supernova Legacy Survey (SNLS, \citealp{Astier06}),
Dark Energy Survey (DES, \citealp{DES16}),
Skymapper Telescope (SMT, \citealp{Scalzo17}),
Wide-Field Infrared Survey Telescope (WFIRST, \citealp{Spergel15,Hounsell17}),
The Large Synoptic Survey Telescope (LSST, \citealp{LSST09}),   %\url{https://github.com/lsst/sims_operations, 
The Asteroid Terrestrial-impact Last Alert System (ATLAS, \citealp{Tonry11}),
Zwicky Transient Facility (ZTF{\footnote{ \url{{https://www.ptf.caltech.edu/page/ztf}}},\citealp{ztf})\footnote{PTF does not have set cadence/depth so is not included here},
and 
All-Sky Automated Survey for Supernovae (ASAS-SN, \citealp{Shappee14}).  

We use the SNANA simulation and analysis package \citep{SNANA} to simulate
each survey using filter transmission functions and a cadence library with a list of observation dates,
where each date includes the observed zero point, sky noise and point spread function (PSF) measured from images.
For SDSS, PS1, SNLS and DES\footnote{Includes Deep and Shallow Fields, numbers listed here are average over all fields}, each cadence library has been created from the actual survey observations,
and therefore includes genuine fluctuations from weather and operational issues.
For LSST, the cadence library is computed\footnote{Observations coadded nightly in \cite{minionsimlibs} from `MINION\_1016'.} from the baseline cadence published by LSST using the Operations Simulator  
\citep{Delgado14},
which uses historical weather data near Cerro Pachon to make realistic estimates of 
observational conditions and cadence.  For WFIRST and SMT, we use the observation libraries 
based on \cite{Hounsell17} and \cite{Scalzo17} respectively.  
For ZTF\footnote{The ZTF simulation done here is for the public survey. Priv. Comm.: Peter Nugent}, ATLAS\footnote{Priv. Comm.: John Tonry} and ASAS-SN\footnote{asas-sn.osu.edu}, we use average quantities for the cadence, zero point, sky noise and PSF.  
The resulting average-cadence libraries do not account for fluctuations from weather, but they are still useful for making forecasts.
Global survey characteristics (depth, cadence, area, duration) are shown in  Table~\ref{tab:hvals} and illustrated in Fig.~\ref{fig:future}. 
There is a dynamic range of 9 magnitudes between the shallowest (ASAS-SN) and deepest (SNLS, WFIRST) surveys, 
and the survey wavelengths extend from the ultraviolet ($u$ band) to the infrared ($F$ band - central wavelength of $1.8$ microns). 
Fig~\ref{fig:future} expresses the cadence as the average gap in time between observations 
with the same filter.  We also show the amount of sky area covered, 
ranging  from 4~deg$^2$ (SNLS) to 18000~deg$^2$ (LSST). 

%ASAS-SN

We simulate KN detections in two steps. The first step is the
trigger simulation, requiring two detections that are separated by at least 30 minutes to reject asteroids.
A detection is characterized by the efficiency vs.
signal-to-noise ratio (S/N), and the efficiency is typically 50\% at $S/N=5$.
The second step is the analysis, which uses the following selection
requirements
designed to reject supernova backgrounds:
\begin{enumerate}
 \item At least two filter bands have at least one observation with $S/N
>5$.
   This requirement is largely redundant with the trigger.
 \item  The time-period when transient measured with $S/N>5$ is less than 25 days (30 days for
WFIRST).
 \item  There is at least one observation within 20 days prior to the
first $S/N>5$ observation.
 \item  There is at least one observation within 20 days after the last
$S/N>5$ observation.
\end{enumerate}
The second requirement explicitly rejects long-lived light curves.
The last two requirements reject events that peak before or after
the survey time window. 
%For example, a SN that explodes well before
%a survey begins can satisfy the second requirement, but would fail
%the second requirement since there are no pre-event observations  with
%a non-detection ($S/N < 5$).

%Various surveys have defined their discoveries and detections differently.  This typically is defined by a number of detections in certain filters. Past analyses (e.g., \citealp{Scolnic14}) defined detections as observations passing a Signal-to-Noise-Ratio (SNR) threshold, though more recent analyses like \cite{Jones16} and \cite{Kessler15} define probabilistic detections given a specific SNR.  Because this study is designed to assess whether KNe can be found with some confidence, we define discovery for every survey as \textit{2 detections with $SNR > {\SNRCUT}$}.

% \tablenotemark{a} 
\begin{table}
\centering
\caption{ Expected number of KNe found in each sample.  }  
\begin{tabular}{l|ccc}
\hline
             &                                           & Survey   & KN Redshift  \\ 
Survey & \# KNe\tablenotemark{a}   &  Years                &  Range \\
\hline
SDSS  & \NKNSDSS\ & \NYSDSS\ & $0.02-0.05$  \\
SNLS  & \NKNSNLS\ & \NYSNLS\  & $0.05-0.20$  \\
PS1    & \NKNPS\ & \NYPS\ & $0.03-0.11$  \\
DES   & \NKNDES\ & \NYDES\ & $0.05-0.20$ \\
ASAS-SN & $<0.001$      &      3          & ---     \\
SMT    & \NKNSMT\ & \NYSMT\ & $0.01-0.01$  \\
ATLAS    & \NKNATLAS\ & \NYATLAS\ & $0.01-0.03$  \\
ZTF        & \NKNZTF\      & \NYZTF\     & $0.01-0.04$  \\
LSST WFD & \NKNLSSTWIDE\ & \NYLSSTWIDE\  & $0.02-0.25$  \\
LSST DDF & \NKNLSSTDEEP\  & \NYLSSTDEEP\ & $0.05-0.25$ \\
WFIRST    & \NKNWFIRST\       & \NYWFIRST\      & $0.1-0.8$          \\
\hline
\end{tabular}
\tablenotetext{1}{Total for entire duration of survey.}
\label{tab:hvals}
\end{table}

% \section{Comparison of Number of Detected KNe}
\section{RESULTS}

% \resizebox{\width}{!}{\input{tables/bnsrates.tex}}\\

% \vspace{.1in}

% \end{table}
%
\begin{figure*}
\begin{center}
\hspace*{-0.2in} 
\scalebox{1.}
{\includegraphics[width=1.0\textwidth]{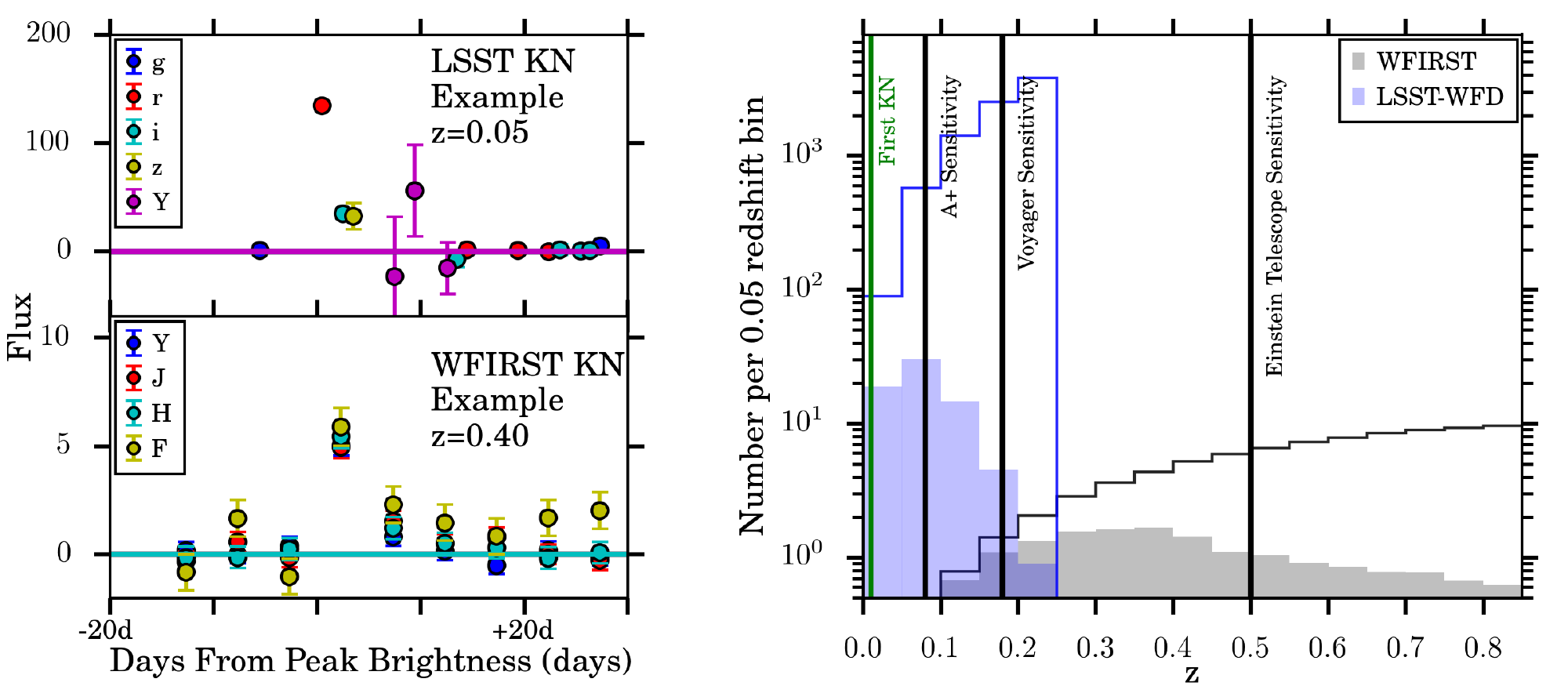}
}
\caption{(Left) Example simulated KN light-curves from LSST and WFIRST that pass our selection requirements.
The vertical axis flux unit is defined such that $mag=27.5-2.5\log(Flux)$.  
(Right) KN redshift distributions for
all events in the survey footprint (solid histogram) and for events 
passing selection requirements (shaded histogram).
Green vertical line shows the KN redshift ($z=\zKN$),
and black vertical lines show the sensitivity of future GW experiments.
}
\label{fig:future}
\end{center}
\end{figure*}

The predicted number of KN detections for each survey
is given in Table 2.  In all of the existing data samples (SDSS, SNLS, PS1, DES, SMT),
the expected number of events is well below unity,
although the expected number is $\sim0.7$ if the KN totals from these 4 surveys  are combined.  Despite the wide variety of area, cadence and depth,
the predicted number of detections in SDSS, SNLS, PS1, DES are all within a factor of $\sim2$.

For future surveys, the estimated rate is larger.  As shown in Table 2, the number of KN discoveries from ATLAS and ZTF is $\sim1-2$ per year, due to their
depth and rapid cadence.
The number of discoveries from LSST WFD is $\sim7$ per year and from LSST DDF is $\sim0.5$ per year.  
Fig.~\ref{fig:future}-left shows a  discovered KN light curve for LSST WFD.  Fig.~\ref{fig:future}-right shows that LSST WFD can discover $<5 \%$ 
of the KNe events in their footprint out to $z=0.25$.

 \begin{figure*}
\begin{center}
\hspace*{-0.2in} 
\scalebox{1.}
{\includegraphics[width=1.0\textwidth]{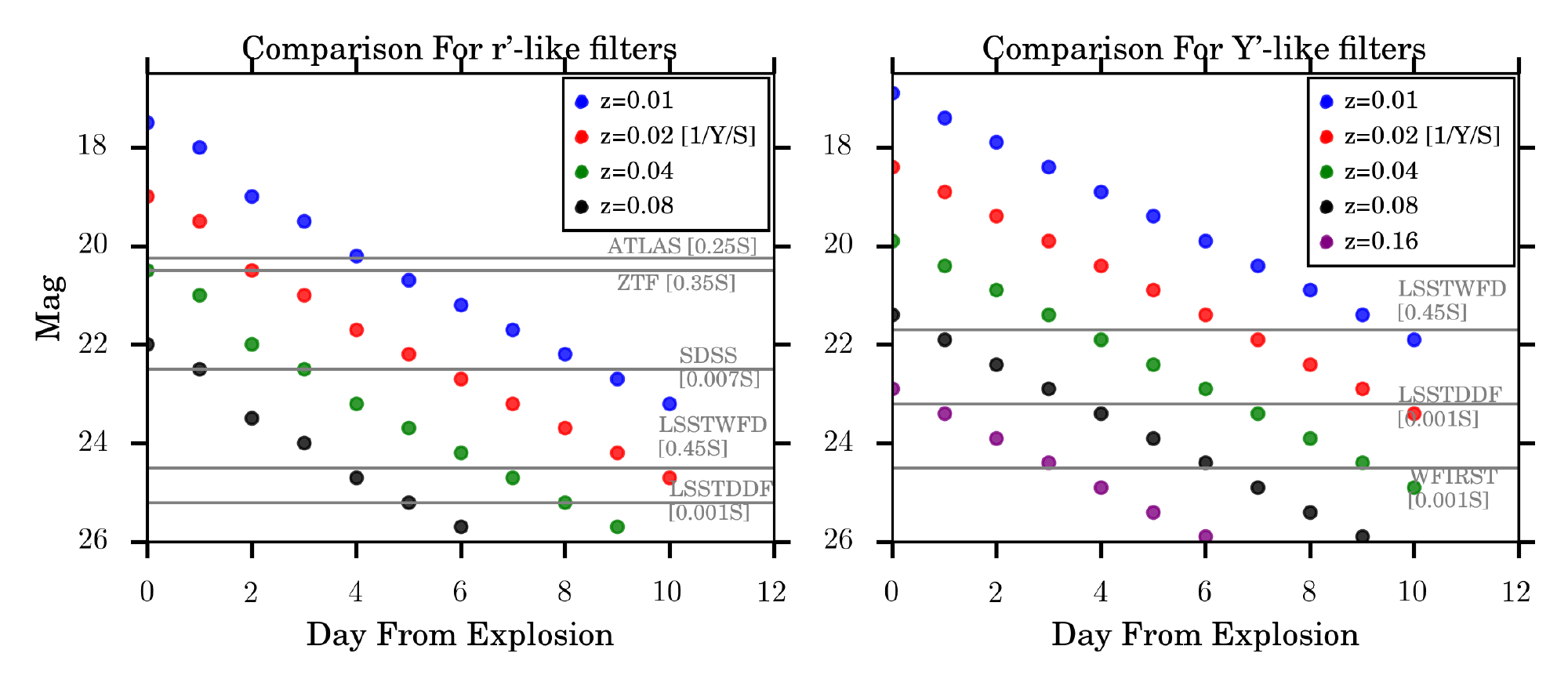}}
\caption{
Synthetic KN light curves at different redshifts (see legend) 
for LSST $r$ band (left) and $Y$ band (right).
Horizontal lines indicate search depth for labelled survey.  
} 
\label{fig:explain}
\end{center}
\end{figure*}

WFIRST has a shorter transient survey duration (2 years),
but still finds as many KNe per season as LSST.  
This KN discovery potential is from a combination of 
depth, medium-sized area, and high red sensitivity.  
We find that the WFIRST efficiency is as high as $\sim 30\%$ in its survey volume.  
Most interestingly, as shown in Fig.~\ref{fig:future}, we see that WFIRST will discover KNe out to 
$z=0.8$.  
Since WFIRST includes observations in the $H$ and $F$ bands, a KN with peak luminosity in the rest-frame $z$ band can still be discovered at $z\sim0.5$ in these red filters.  

To illustrate the interplay between depth, rate, and sky coverage, 
we show in Fig.~\ref{fig:explain} the $r$ and $Y$ detection limits of multiple surveys 
overlaid on our KN lightcurve as it would appear at 
discrete redshifts.
 
\subsection{ Background Contamination From Supernovae}

With {\NKNLSSTWIDE} KN events expected for the LSST WFD survey,
we now switch to simulating the background from supernova (SNe).  We include Type Ia SNe (SNIa) based on the 
SALT2-II spectral model \citep{Betoule14},
and core collapse (CC) SNe based
on a library of 43 templates \citep{Kessler10}. % end blue

%   [NOTE: I think we already commented on extensive SNANA use in the introduction]
%Over the past decade, these SN simulations with SNANA
%have been extensively used to simulate distance biases
%in cosmology analyses (e.g., \citealp{Jones16,Scolnic14,Betoule14}).

\newcommand{\Pfit}{P_{\rm fit}}

In addition to the trigger and analysis requirements in \S 3,
we use the PSNID fitting program \citep{Sako08} to select KN-like
objects via light curve template matching. The templates include
SNIa, Type II SNe, Type Ib/c SNe, and our KN event.
For a given simulated CC template, the corresponding template is
removed from the PSNID fit so that we don't match a
simulated CC template to itself. 
%While the trigger and analysis requirements rely mainly on the redder bands,
This PSNID analysis uses bands at all wavelengths, and thus even a KN flux limit 
in the bluer bands add useful information.

For the full 10 year survey, we generated nearly 200 million SNe (16\% Ia, 84\% CC),
and find that 9 events are identified as a KN by PSNID.
However, only 2 of these events have a reasonable fit-probability,
$\Pfit > 0.001$, where $\Pfit$ is computed from the $\chi^2$ and
number of degrees of freedom.  
This background is $2.9\%$ of the number of KNe detected.

\section{Discussion}
 
None of the surveys discussed here have been optimized to find KNe, so the KN yields are expected to be low.  SDSS, PS1, SNLS, LSST DDF, and WFIRST are all partially optimized for measurement of SNIa light-curves, which have typical durations of 60 days. 
While we expect $\sim1$ KNe in past datasets, 
we note that it is unlikely to find such an event in light-curve catalogs. 
Instead, a search for KNe in old data requires a re-analysis of all single-epoch detections to make less strict trigger cuts than 
those applied during past surveys.  
 As improved volumetric KN rate estimates become available, 
all of our KN predictions can be re-scaled.

We have performed a preliminary study of SN background, 
and while this small (3\%) background  is encouraging, we note
a few caveats that require further study.  
First, our simulations do not include potential contaminants from rare 
SN types, moving objects (asteroids),
and non-SN transients such as orphan afterglows of 
GRBs (e.g., \citealp{Singer13}) and M-dwarf flares (e.g., \citealp{Hawley14}).

The second caveat is that we have implicitly assumed that all KN are the same, 
which is very unlikly to be correct.  
Ideally, our single KN template should be expanded
to accept a wide range of KNe, perhaps with the aid of theoretical models such as
\cite{Barnes2016}.  However, the challenge is to keep the SN backgrounds low while accepting a broader class of KN events.

Another caveat is that we have used the full end-of-survey light curves,
but to get crucial follow-up observations with other instruments, KN events need to be
efficiently identified within a few days of the merger event. Partial light curve studies
will be needed to optimize KN target selection. 

The final caveat is related to the KN host galaxies.
In a recent search of DES-SN data (without a GW trigger), \cite{Doctor17} found that 
image-subtraction artifacts increase the flux scatter well beyond what is expected
from Poisson noise, and thus reduces the search sensitivity by a factor of 3 if all KNe 
occur inside their host galaxy.  For KN events like this one, the event is well away from the galaxy center, as is expected for the majority of short GRB \citep{Fong13}.  Therefore, image-subtraction artifacts are likely to be a subdominant issue, though the impact on expected KN should still be quantified.

One of the most interesting findings of this analysis is the ability for WFIRST to discover high-redshift KNe.
This is particularly exciting because  it would probe the cosmic
history of NS-mergers.  Furthermore, it could provide an absolute distance scale to $z\sim0.5$, which could be the first absolute distance measurement in between the 
local and CMB Hubble constant measurements.
What's also illuminating is that WFIRST may detect KNe at higher redshift than the sensitivity of future GW missions.   Chen et al. (in prep.), based on methodology from \cite{Chen17}, estimates the sensitivity of next generation gravitational wave detectors and we mark these sensitivities on Fig.~\ref{fig:future}.  
We find that the LIGO upgrade A\texttt{+}~design,
the future detector LIGO Voyager, and
the planned Einstein Telescope 
all have sensitivity to GW triggers below the depth of WFIRST to KN events.  Furthermore, theoretical models consistent with \cite{CowperthwaiteLC} suggest that the
blue component  depends on viewing angle, while the red component is
isotropic. The IR capability of WFIRST may therefore have the additional advantage
of better sensitivity to all viewing angles. There is an ongoing effort to design a joint GW and WFIRST program, 
called \textit{GWFIRST}, optimized for NIR follow-up of GW detections.

Lastly, this analysis only looks at survey detections
without a GW trigger,
whereas the most likely mode for most telescopes will be follow-up of announced GW events.  
With estimates of area, cadence and observing conditions,
all of the simulation tools used here can be used to optimize follow-up strategies.

\section{Conclusion}
 We have used simulations to predict the number of KNe that can be found in 
past, present and future datasets. 
The simulation uses a KN model  that matches our DECam light curve data,
and for each survey it uses realistic observation histories.  
We find that the expected number of events for every past survey is 
$\sim0-0.3$ due to the small area, shallow depth, or sparse cadence, though combined can be up to $\sim1$ event.  For future surveys like LSST and WFIRST, we expect tens of KN discoveries.  In particular, we find that WFIRST can find KNe at redshifts past planned GW sensitivities of future projects, opening up new possibilities of cosmological KN and NS science.

\acknowledgements
Funding for the DES Projects has been provided by the DOE and NSF(USA), MEC/MICINN/MINECO(Spain), STFC(UK), HEFCE(UK). NCSA(UIUC), KICP(U. Chicago), CCAPP(Ohio State),MIFPA(Texas A\&M), CNPQ, FAPERJ, FINEP (Brazil), DFG(Germany) and the Collaborating Institutions in the Dark Energy Survey. The Collaborating Institutions are Argonne Lab, UC Santa Cruz, University of Cambridge, CIEMAT-Madrid, University of Chicago, University College London, DES Brazil Consortium, University of Edinburgh, ETH Zürich, Fermilab, University of Illinois, ICE (IEEC-CSIC), IFAE Barcelona, Lawrence Berkeley Lab, LMU München and the associated Excellence Cluster Universe, University of Michigan, NOAO, University of Nottingham, Ohio State University, University of Pennsylvania, University of Portsmouth, SLAC National Lab, Stanford University, University of Sussex, Texas A\&M University, and the OzDES Membership Consortium. Based in part on observations at Cerro Tololo Inter-American Observatory, National Optical Astronomy Observatory, which is operated by the Association of Universities for Research in Astronomy (AURA) under a cooperative agreement with the National Science Foundation.

The DES Data Management System is supported by the NSF under Grant Numbers AST-1138766 and AST-1536171.  This work was supported in part by the Kavli Institute
for Cosmological Physics at the University of Chicago
through grant NSF PHY-1125897 and an endowment
from the Kavli Foundation and its founder Fred Kavli.
We gratefully acknowledge support from NASA grant
14-WPS14-0048. 
D.S. is supported by NASA through
Hubble Fellowship grant HST-HF2-51383.001 awarded
by the Space Telescope Science Institute, which is operated
by the Association of Universities for Research
in Astronomy, Inc., for NASA, under contract NAS 5-
26555. This analysis was done using the Midway-RCC computing cluster at University of Chicago.  The Berger Time-Domain Group at Harvard is supported
in part by the NSF through grants AST-1411763 and AST-
1714498, and by NASA through grants NNX15AE50G and
NNX16AC22G. The UCSC group is supported in part by NSF grant AST--1518052, the Gordon \& Betty Moore Foundation, the Heising-Simons Foundation, generous donations from many individuals through a UCSC Giving Day grant, and from fellowships from the Alfred P.\ Sloan Foundation and the David and Lucile Packard Foundation to R.J.F.  RB acknowledges partial support from the Washington Research Foundation Fund for Innovation in Data-Intensive Discovery and the Moore/Sloan Data Science Environments Project at the University of Washington.

%\bibliographystyle{apj.bst}
%\bibliography{bib}

\end{document}